\documentclass[manuscript]{aastex}

\usepackage{hyperref}
\hypersetup{colorlinks=true,
            citecolor=blue,
            linkcolor=blue}

\def\vv#1{\mathbf{#1}}
\def\be {\begin{equation}}
\def\ee {\end{equation}}
\def\ve {\varepsilon}
\def\vr {\vv{r}}
\def\vi {\vv{i}}
\def\vj {\vv{j}}
\def\vk {\vv{k}}
\def\TI {{\bf \cal I}}
\def\ii {\rm i}
\def\M {m_\star}
\def\om {\Omega}
\def\lv {\psi}
\def\lw {\omega}

\def\crm{\cr\noalign{\medskip}}

\def \llabel#1{\label{#1}}

\def\epsl{Earth Planet.\ Sci.\ Lett. }

\shorttitle{On the equilibrium figure of close-in planets and satellites}
\shortauthors{Correia \& Rodr\'iguez}

\begin{document}


\title{On the equilibrium figure of close-in planets and satellites}


\author{Alexandre C.M. Correia}
\affil{Departamento de F\'isica, I3N, Universidade de Aveiro, Campus de
Santiago, 3810-193 Aveiro, Portugal;}
\affil{ASD, IMCCE-CNRS UMR8028, Observatoire de Paris, UPMC,
77 Av. Denfert-Rochereau, 75014 Paris, France}

\and

\author{Adri\'an Rodr\'iguez}
\affil{Insituto de Astronomia, Geof\'isica e Ci\^encias Atmosf\'ericas, IAG-USP, Rua do Mat\~ao 1226, 05508-090 S\~ao Paulo, Brazil}


\begin{abstract}
Many exoplanets have been observed close to their parent stars with orbital periods of a few days.
As for the major satellites of the Jovian planets, the figure of these planets is expected to be strongly shaped by tidal forces.
However, contrarily to Solar System satellites, exoplanets may present high values for the obliquity and eccentricity due to planetary perturbations, and may also be captured in spin-orbit resonances different from the synchronous one.
Here we give a general formulation of the equilibrium figure of those bodies, that makes no particular assumption on the spin and/or orbital configurations.
The gravity field coefficients computed here are well suited for describing the figure evolution of a body whose spin and orbit undergo substantial variations in time.
\end{abstract}





\section{Introduction}

All the main satellites of the giant planets in the Solar System are in synchronous rotation, and locked in a nearly zero-obliquity Cassini state; their orbits are nearly circular, with orbital periods less than $16$~days, and lie in the equatorial plane of the central planet\footnote{http://ssd.jpl.nasa.gov/}.
These configurations are the result of the long-term evolution of the spin and orbits due to tidal forces raised by the central planet \citep[e.g.][]{Correia_2009}.
The influence of the tidal deformation on the shape of the satellites is also appreciable, because the satellites' figures are supposed to be in hydrostatic equilibrium with the tidal and centrifugal potentials \citep[e.g.][]{Schubert_etal_2004}. 
This assumption results from the fact that their interiors are either hot (Io and Europa) or plastic (icy satellites, e.g., Ganymede, Callisto, and Titan). 

About 40\% of the presently known exoplanets are found in orbits with periods below 16~days, and at least more than 25\% of them are not alone in their systems\footnote{http://exoplanet.eu/}.
As for the satellites above, we can assume that these planets acquire an equilibrium shape dictated by its internal gravity and a perturbing potential. 
However, due to planetary perturbations from additional companions, close-in exoplanets may present non-zero obliquity \citep[e.g.][]{Laskar_Robutel_1993,Levrard_etal_2007} or eccentric orbits \citep[e.g.][]{Correia_Laskar_2004,Mardling_2007}. 
In addition, they may also be captured in spin-orbit resonances different from the synchronous one \citep[e.g.][]{Goldreich_Peale_1966,Rodriguez_etal_2012}.

Previous works have shown that non-synchronous satellites present a different figure \citep{Zharkov_Leontev_1989, Giampieri_2004}.
In addition, if the orbit has some eccentricity, the low order gravity coefficients can be separated into static and periodic components (involving different time-scales), 
and only the static part contributes to the permanent deformation \citep{Rappaport_etal_1997, Giampieri_2004}.
Therefore, since exoplanets may evolve in very eccentric orbits, their figures are most likely different from what we observe in Solar System satellites.

In this Letter we generalize previous studies to any eccentricity value.
We also include the effect of an arbitrary obliquity, that may change completely the gravity field coefficients. 
We adopt a vectorial formalism, which is different from the traditional expansion of the gravitational potential in spherical harmonics.

\section{The gravitational potential}

The gravitational potential of a body of mass $m$ at a generic point $P$ is given by:
\be 
V (\vr) = - G \int{ \frac{d m}{| \vr - \vr' |}} 
\llabel{121026a} 
\ee
where $G$ is the gravitational constant, $ \vr $ is the position of $ P $, and $ \vr' $ the position of a mass element $ d m $ with respect to center of mass of the body. 
Assuming $ r' \ll  r $, we can develop the potential limited to the second order, which gives  \cite[e.g.][]{Tisserand_1891,Smart_1953}:
\be 
V (\vr) = - \frac{G m}{r} + \frac{3 G}{2 r^3} \left( \hat \vr \cdot \TI \cdot \hat \vr - \frac{1}{3} tr(\TI) \right) \ , \llabel{121026b}
\ee
where $\hat \vr = \vr / r $,
\be
\TI = 
\left[\begin{array}{rrr} 
I_{11}&  I_{12}& I_{13} \crm
I_{12}&  I_{22}& I_{23} \crm
I_{13}&  I_{23}& I_{33} 
\end{array}\right] 
\label{121026c}
\ee
is the itertia tensor, and $tr(\TI) = I_{11} + I_{22} + I_{33} $ its trace.
We chose as reference the non-inertial frame ($\vi,\vj,\vk$) fixed to the planet, where $\vk$ is the direction of the spin axis (Fig.\,\ref{fig1}).

\begin{figure}
\centering
\includegraphics*[width=16.cm]{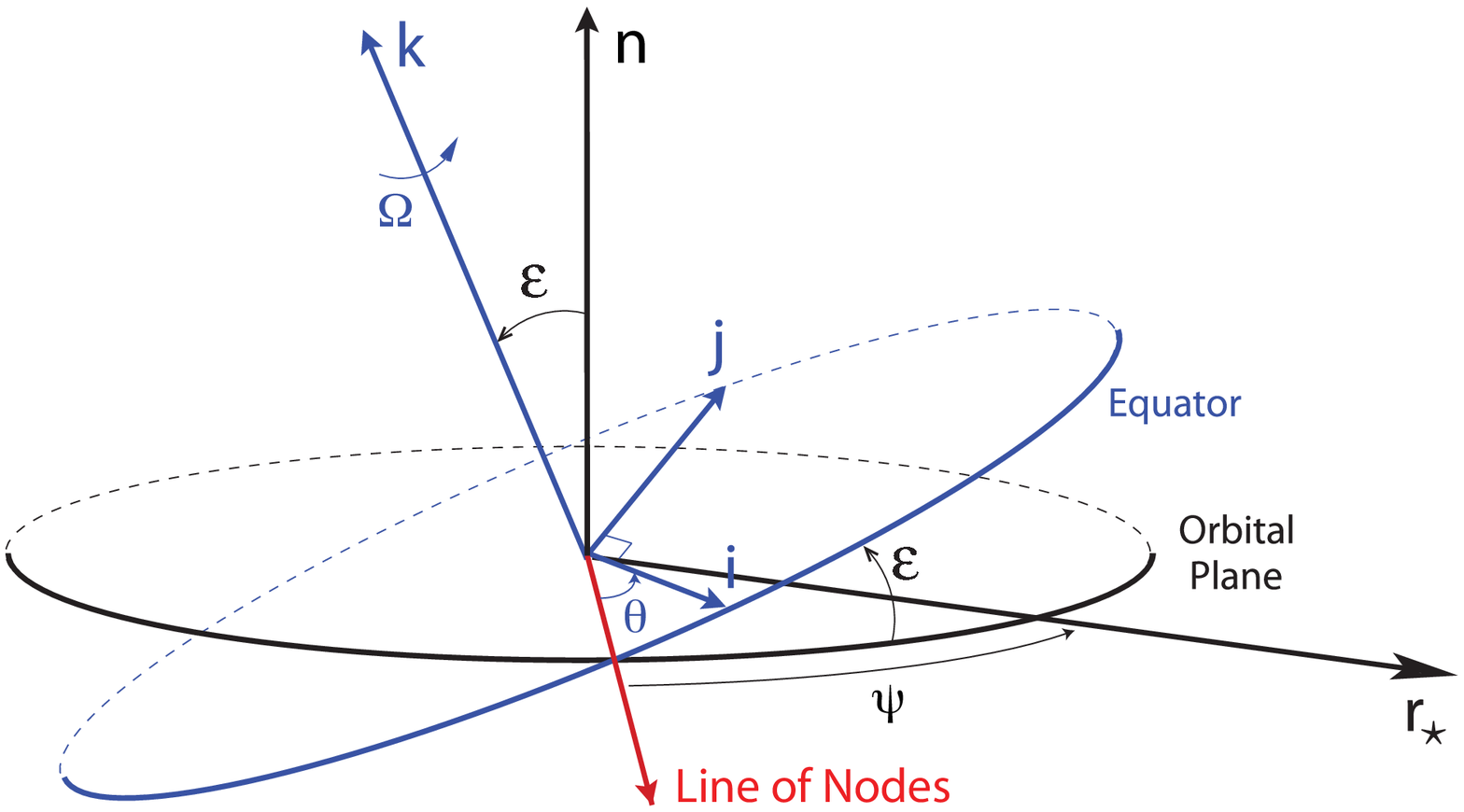}
\caption{Reference frames for the definition of the reference angles. 
  $ \vv{n} $ is the normal vector to the orbital plane of the perturbing body at $
  \vr_\star $. The reference frame ($\vi, \vj, \vk$) is fixed with respect to the 
  planet's figure, $\vk$ being the rotation axis, that is tilted by an angle $\ve$ (the obliquity) with respect to $\vv{n}$.
  The rotation angle, $\theta$, and the position of the perturber, $\lv$, are measured from the line of nodes between the equatorial and orbital planes, along their respective planes.
  $\Omega $ is the rotation rate.
  \llabel{fig1} }
\end{figure}

The above potential is also often expressed  alternatively using the gravity field coefficients $J_2$, $C_{2n}$, $S_{2n}$, as
\begin{eqnarray}
V (\vr) &=& - \frac{G m}{r} - \frac{G m R^2}{r^3} \left[  \phantom{\frac{}{}} C_{20} \, P_2 (\vk \cdot \hat \vr)  
\right. \crm &+& \left.
C_{21} \, 3 (\vi \cdot \hat \vr) (\vk \cdot \hat \vr) + S_{21} \, 3 (\vj \cdot \hat \vr) (\vk \cdot \hat \vr) 
\right. \crm &+& \left.
C_{22} \, 3 ((\vi \cdot \hat \vr)^2 - (\vj \cdot \hat \vr)^2)+ S_{22} \, 6 (\vi \cdot \hat \vr) (\vj \cdot \hat \vr)
 \phantom{\frac{}{}} \right] \ , \llabel{121102b}
\end{eqnarray}
where $P_2(x)=(3x^2-1)/2$, and
\be
J_2= -C_{20} = \frac{I_{33}-\frac{1}{2} (I_{11}+I_{22})}{m R^2}
\ , \llabel{121102c}
\ee
\be
C_{21}= -\frac{I_{13}}{m R^2}
\ , \quad 
S_{21}= -\frac{I_{23}}{m R^2}
\ , \llabel{121102d}
\ee
\be
C_{22}= \frac{I_{22}-I_{11}}{4 m R^2}
\ , \quad 
S_{22}= -\frac{I_{12}}{2 m R^2}
\ . \llabel{121102e}
\ee

\section{The perturbing potential}

The mass distribution in the body, characterized by the inertia tensor (Eq.\,\ref{121026c}), is a result of the self gravity, but also of the body's response to any perturbing potential.

For a planet with rotation rate $\vv{\om} = \om \, \vk$, a mass element $d m$ is also subject to the centrifugal potential \citep[e.g.][]{Goldstein_1950}
\be
V_c (\vr') = - \frac{1}{2} (\vv{\om} \times \vr')^2 = - \frac{1}{2} \om^2 r'^2 \left( 1 - (\vk \cdot \hat \vr')^2 \right) \ . \llabel{121026d}
\ee

In addition, since the planet moves around a central star of mass $\M$, we also need to consider the tidal potential \citep[e.g.][]{Lambeck_1980}
\be
V_t (\vr', \vr_\star) = -\frac{G \M}{r_\star} \left(\frac{r'}{r_\star}\right)^2 P_2 (\hat \vr' \cdot \hat \vr_\star) \ , \llabel{121026e}
\ee
where  $\vr_\star$ is the position of the star with respect to the planet center of mass. 
In the frame of the planet, we can express $\hat \vr_\star$ as \citep[e.g.][]{Correia_2006}:
\be
\frac{\vr_\star}{r_\star} = 
\left[\begin{array}{c} 
\cos \ve \sin \lv \sin \theta +\cos \lv \cos \theta \crm
\cos \ve \sin \lv \cos \theta -\cos \lv \sin \theta  \crm
-\sin \ve \sin \lv 
\end{array}\right] \ ,
\label{121026z}
\ee
where $ \ve $ is the obliquity of the planet (the angle between the spin axis and the normal to the orbit), $\theta$ is the rotation angle, $\lv = \lw + \nu $ is the angle between the line of nodes and the direction of the star,  $\lw$ is the argument of the periapse, and $\nu$ is the true anomaly (Fig.\,\ref{fig1}).

The resulting perturbing potential is then given by
\be
V_p (\vr', \vr_\star) = V_c (\vr') + V_t (\vr', \vr_\star)  \llabel{121026f} \ ,
\ee
that can also be rearranged as
\be
V_p (\vr', \vr_\star) =  \left( \frac{G \M}{r_\star^3} -  \om^2 \right) \frac{r'^2}{2}
+\frac{3 G}{2 r'^3} ( \hat \vr' \cdot \TI_p \cdot \hat \vr' )  \llabel{121030a} \ ,
\ee
where
\be
\TI_p = \TI_c + \TI_t  \llabel{121030b} \ ,
\ee
with
\be
\frac{\TI_c}{m r'^2} = \frac{\om^2 r'^3}{3 G m}
\left[\begin{array}{rrr} 
0& 0& 0 \crm
0& 0& 0 \crm
0& 0& 1 
\end{array}\right] 
\llabel{121030c} \ ,
\ee
and
\be
\frac{\TI_t}{m r'^2} = -  \frac{\M}{m}  \left( \frac{r'}{r_\star} \right)^3 \left[  \frac{\vr_\star}{r_\star} \right]^T \left[  \frac{\vr_\star}{r_\star} \right] 
\llabel{121030d} \ .
\ee
The last term in the above equation is directly obtained from expression (\ref{121026z}), where $^T$ denotes the transpose.

\section{Equilibrium figure}
\llabel{eqfig}

The perturbing potential (Eq.\,\ref{121026f}) acting on the planet, deforms it, and modifies the external gravitational potential (Eq.\,\ref{121026a}).
A most convenient way of defining this deformation is through the Love number approach \citep[e.g.][]{Love_1927, Peltier_1974}, in which the body's response is evaluated in the frequency domain.
As long as the distortions are small, we can simplify the problem by ignoring the small
interaction terms between the centrifugal and tidal potentials \citep{Zharkov_Trubitsyn_1978}.
The additional gravitational potential, $\Delta V (\vr)$, due to the deformation of the planet in response to the perturbing potential, is then given at the planet's surface ($\vr = \vv{R}$) 
by \citep[e.g.][]{Lambeck_1980}:
\be
\Delta V(\vv{R}) = k_2 V_p (\vv{R}, \vr_\star) 
\ , \llabel{121102h}
\ee
where $k_2$ is the second Love number for potential.
In general, $k_2$ is a complex number, which depends on the frequency  $\sigma$ of the perturbation.
$|k_2|$ gives the amplitude of the tide, while the imaginary part gives the phase lag between the perturbation and the maximal deformation.

If we assume that the planet behaves like a Maxwell body\footnote{A Maxwell body
behaves like an elastic body over short time scales, but flows like a fluid over long periods of time. It is characterized by a homogenous rigidity $ \mu $ and viscosity $ \upsilon $.}
with homogeneous density $ \rho $ we have \citep{Henning_etal_2009}:
\be
k_2 (\sigma) = k_f   \frac{1 + \ii \, \sigma \tau_a} 
{1 + \ii \, \sigma \tau_b }
\ , \llabel{V5a} 
\ee
where $k_f$ is the fluid second Love number, $\tau_a = \upsilon / \mu$ is the relaxation time, and
\be
\frac{\tau_b}{\tau_a} = \left(1 + \frac{19 \mu R}{2 G m \rho} \right) = \frac{k_f}{k_2(\infty)} \llabel{121112b} \ .
\ee 
$k_f$ is constant for a given body and corresponds to its maximal deformation
(for a static perturbation $\sigma = 0$,  and hence $k_2 = k_f$).
For an homogeneous sphere $k_f = 3/2$, but more generally $k_f$ can
can be obtained from the Darwin-Radau equation \citep[e.g.][]{Jeffreys_1976}:
\be
\frac{I_{33}}{m R^2} = \frac{2}{3} \left(1 - \frac{2}{5} \sqrt{\frac{4 - k_f}{1+ k_f}} \right)  \ . \llabel{121112a}
\ee

In Figure\,\ref{Fk2} we plot the dependence of $|k_2|$ 
with the tidal frequency for the Earth.
Since static and elastic stresses involve very different time scales, there is no conflict between the two types of responses (Fig.\,\ref{Fk2}). 
We can thus assume a static response for the centrifuge potential (Eq.\,\ref{121026d}) and an elastic one for the tidal potential (Eq.\,\ref{121026e}), which depends on the varying position of the star with respect to a point at the planet's surface.

\begin{figure}
\centering
\includegraphics*[width=16.cm]{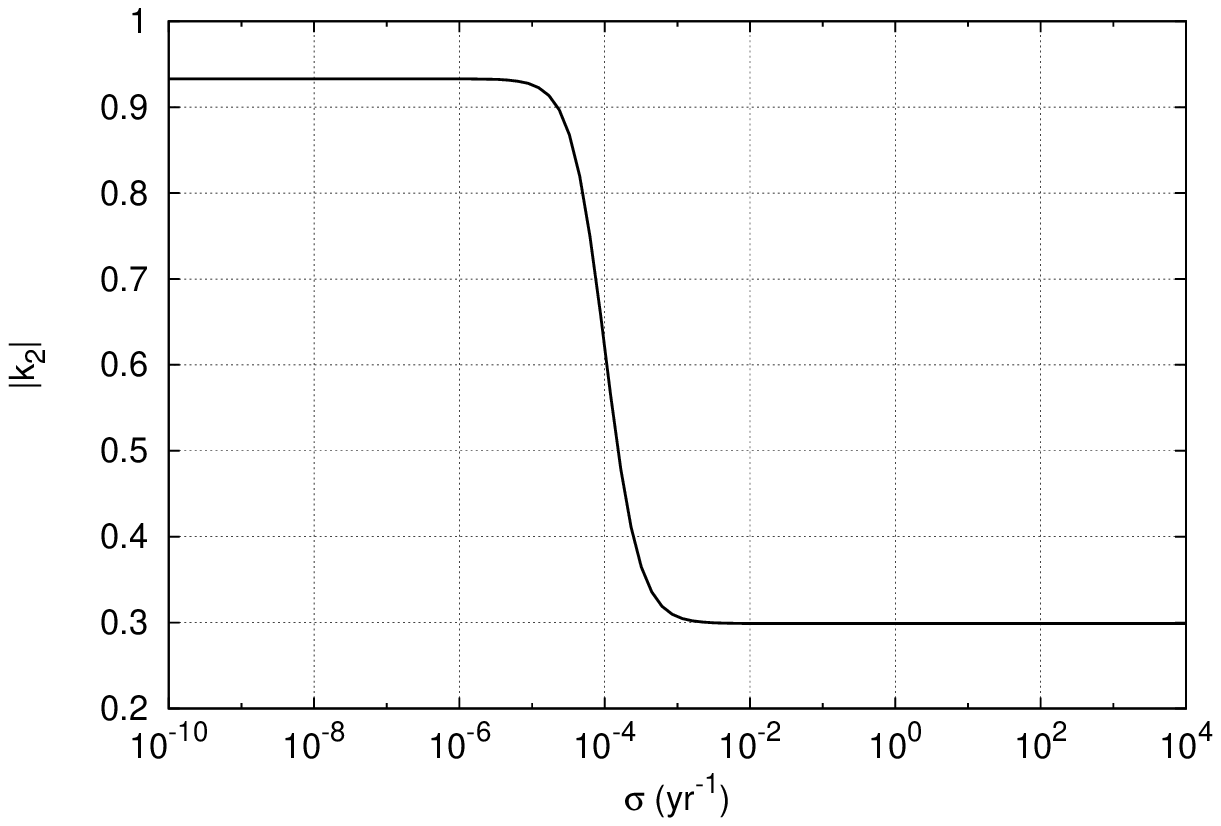}
\caption{Dependence of the amplitude $| k_2 |$ 
with the tidal frequency $\sigma$ for the Earth. We use a visco-elastic model (Eq.\,\ref{V5a}) with $\upsilon \sim 10^{22} \, \mathrm{kg\,m^{-1} s^{-1}}$ \citep{Cizkova_etal_2012}, $k_2 = 0.299$, and $k_f = 0.933 $ \citep{Yoder_1995}.
  \llabel{Fk2} }
\end{figure}

Combining expressions (\ref{121026b}, \ref{121030a}, and \ref{121102h}) we thus have
\be
\TI =  \TI_0 + k_2 \, \TI_p (R)
\ , \llabel{121103a}
\ee
where $\TI_0 \propto \mathbb{I}$ corresponds to the moment of inertia with spherical symmetry, $\mathbb{I}$ being the identity matrix.
It becomes then straightforward to compute the gravity coefficients (Eqs.\,\ref{121102c}$-$\ref{121102e}), for which $\TI_0$ is irrelevant. 

For the centrifuge contribution, all terms in $\TI_c$ are zero except $I_{33}^c$ (Eq.\,\ref{121030c}), thus the only non-zero gravity coefficient is $J_2$:
\be
J_2^c = k_2 (0) \frac{I_{33}^c (R)}{m R^2} = k_f \frac{\om^2 R^3}{3 G m}
\llabel{121102i} \ .
\ee

For the tidal contribution, the inertia matrix is permanently modified (Eq.\,\ref{121030d}), meaning that all gravity coefficients are non-zero.
The global contribution to the gravity coefficients is given by the sum of the centrifuge and tidal contribution (Eq.\,\ref{121030b}), i.e.,
\be
J_2 = J_2^c + J_2^t \ , \llabel{121105a}
\ee
with
\be
J_2^t =   \frac{k_2}{2} \frac{\M}{m} \left(\frac{R}{r_\star}\right)^3 \left[ 1-\frac{3}{2} \sin^2 \ve \, (1-\cos 2\lv) \right] \llabel{121105b} \ ,
\ee
while for the remaining coefficients, the tidal contribution is the total one:
\begin{eqnarray}
C_{21}&=& - \frac{k_2}{2} \frac{\M}{m} \left(\frac{R}{r_\star}\right)^3 \sin \ve \left[  
\phantom{\frac{}{}} \cos \ve  \sin \theta 
\right. \crm &-& \left.
\cos^2 \frac{\ve}{2} \sin (\theta-2\lv)  + \sin^2 \frac{\ve}{2} \sin (\theta+2\lv)
 \phantom{\frac{}{}} \right] \ , \llabel{121105c} 
\end{eqnarray}
\begin{eqnarray}
S_{21}&=& - \frac{k_2}{2} \frac{\M}{m} \left(\frac{R}{r_\star}\right)^3 \sin \ve \left[  
\phantom{\frac{}{}} \cos \ve \cos \theta 
\right. \crm &-& \left.
\cos^2 \frac{\ve}{2}  \cos (\theta-2\lv)  + \sin^2 \frac{\ve}{2}  \cos (\theta+2\lv)
 \phantom{\frac{}{}} \right] \ , \llabel{121105d} 
\end{eqnarray}
\begin{eqnarray}
C_{22}&=& \frac{k_2}{8} \frac{\M}{m} \left(\frac{R}{r_\star}\right)^3 \left[  
\phantom{\frac{}{}} \sin^2 \ve \, \cos 2 \theta
\right. \crm &+& \left.
2 \cos^4 \frac{\ve}{2} \cos 2 (\theta-\lv) + 2 \sin^4 \frac{\ve}{2} \cos 2 (\theta+\lv)
\phantom{\frac{}{}} \right] \ , \llabel{121105e} 
\end{eqnarray}
\begin{eqnarray}
S_{22}&=& - \frac{k_2}{8} \frac{\M}{m} \left(\frac{R}{r_\star}\right)^3 \left[  
\phantom{\frac{}{}} \sin^2 \ve \,  \sin 2 \theta
\right. \crm &+& \left.
2 \cos^4 \frac{\ve}{2} \sin 2 (\theta-\lv) + 2 \sin^4 \frac{\ve}{2} \sin 2 (\theta+\lv)
\phantom{\frac{}{}} \right] \ . \llabel{121105f} 
\end{eqnarray}
Note that, unless the planet evolves in a circular orbit, we cannot replace $k_2$ by $k_f$ in expression (\ref{121105b}) for the terms independent of $\lv$, because the distance to the star also varies with $\lv$ as \citep[e.g.][]{Murray_Dermott_1999}:
$ r_\star = a (1 - e^2) / (1 + e \cos \nu) $, 
where $a$ and $e$ are the semi-major axis and the eccentricity of the orbit, respectively. 

\section{Permanent deformation}

In the previous section 
we assumed for the tidal perturbation an elastic response, but this is not completely true for all harmonics. 
Indeed, although the position of the star with respect to the planet surface may not be constant, when we average its motion over one orbital period, some perturbations do not average to zero, and the planet can assume a different permanent figure.
In order to identify the static harmonics, we need to perform an expansion of the true anomaly $\nu$ in series of the eccentricity $e$ and mean anomaly $M$: 

\be 
\frac{\mathrm{e}^{\ii k \theta}}{r_\star^3} = \frac{1}{a^3} \sum^{+\infty}_{q=-\infty} G(q,e) \mathrm{e}^{\ii (k \theta - 2 q M)} \ , \llabel{061120ga}
\ee
and  
\be 
\frac{\mathrm{e}^{\ii (k \theta - 2 \nu)}}{r_\star^3} = \frac{1}{a^3} \sum^{+\infty}_{q=-\infty} H(q,e) \mathrm{e}^{\ii (k \theta - 2 q M)} \ , \llabel{061120gb}
\ee  
where the functions $G(q,e)$ and $H(q,e)$ are power
series in $e$ (Tab.\,\ref{TAB1}), and $k$ and $2 q$ are integers\footnote{We have retained the use of semi-integers for $q$ for a better  comparison with previous works.}.
Static perturbations thus correspond to frequencies for which $\sigma = k \dot \theta - 2 q \dot M = 0$.
They can occur whenever $k = q = 0$, or if there is a commensurability between the rotation rate and the orbital motion (spin-orbit resonances).

\begin{table}
\begin{center}
\caption{Hansen coefficients $G(q,e)$ and $H(q,e)$ to  $e^4$.  \llabel{TAB1} }
\begin{tabular}{|r|r|r|} \hline 
$ q $ &  $ G ( q , e ) $ & $ H ( q , e ) $ \\ \hline \hline
$ -1/1$  & $ \frac{9}{4} e^2  +  \frac{7}{4} e^4 $ &  $ \frac{1}{24} e^4 $ \\ 
$ -1/2$  &  $ \frac{3}{2} e + \frac{27}{16} e^3 $ & $ \frac{1}{48} e^3 $  \\ 
$  0/1 $ & $ 1 +  \frac{3}{2} e^2  +  \frac{15}{8} e^4 $ & $ 0 $ \\ 
$  1/2 $ & $ \frac{3}{2} e + \frac{27}{16} e^3 $ & $ -  \frac{1}{2} e  + \frac{1}{16} e^3 $  \\ 
$  1/1 $ & $ \frac{9}{4} e^2 +  \frac{7}{4} e^4 $ & $ 1 - \frac{5}{2} e^2 + \frac{13}{16} e^4 $ \\ 
$  3/2 $ & $ \frac{53}{16} e^3 $ & $ \frac{7}{2} e - \frac{123}{16} e^3 $ \\ 
$  2/1 $ & $ \frac{77}{16} e^4 $ & $ \frac{17}{2} e^2 - \frac{115}{6} e^4 $ \\ 
$  5/2 $ & & $ \frac{845}{48} e^3 $  \\ 
$  3/1 $ & & $ \frac{533}{16} e^4 $ \\ \hline
\end{tabular} 
\end{center}
The exact expression of these coefficients is given by $G(q,e) =
\frac{1}{\pi} \int_0^\pi \left( \frac{a}{r} \right)^3 \exp(\ii \, 2 q M) \, d M
$ and $H(q,e) = \frac{1}{\pi} \int_0^\pi \left( \frac{a}{r} \right)^3
\exp(\ii \, 2 \nu) \exp(\ii \, 2 q M) \, d M $.
\end{table}

\subsection{Fast rotating planets}

For an arbitrary rotation rate (such as the rotation of the Earth), only terms with $k = q = 0$ will contribute to a permanent deformation of the planet.
By replacing equations (\ref{061120ga}) and (\ref{061120gb}) in the expressions of the gravity coefficients (Eqs.\,\ref{121105b}$-$\ref{121105f}), and averaging over the rotation angle $\theta$ and the mean anomaly $M$, we get that all gravity coefficients become zero, except $J_2^t$:
\be
< J_2^t > = \frac{k_f}{2} \frac{\M}{m} \left(\frac{R}{a} \right)^3 G(0,e) \left[1-\frac{3}{2} \sin^2 \ve \right] \llabel{121105z} \ ,
\ee
where $G(0,e)=(1-e^2)^{-3/2}$.
Although for eccentric orbits the distance to the star constantly varies with $\lv$, 
there is a permanent perturbation along the direction of the two bodies whose average is not zero.

\subsection{Spin-orbit resonances}

When the spin is tidally evolved, it may be captured in a spin-orbit resonance, for which $ \om = p n $ \citep[e.g.][]{Goldreich_Peale_1966,Correia_Laskar_2009}.
The most common outcome is the synchronous resonance ($ p = 1 $), observed in the Solar System for the majority of the main satellites.
However, non-synchronous configurations are also possible, as it is the case of Mercury ($ p = 3/2 $) \citep{Colombo_1965}.

For synchronous rotation, the planet acquires a permanent deformation along the direction that always point to the perturber \citep[e.g.][]{Ferraz-Mello_etal_2008}.
For non-synchronous resonances, in average, one direction always points to the perturber at the periapse \citep{Goldreich_Peale_1966}.
Thus, bodies with some rigidity can also acquire a permanent deformation along this direction. 

In order to obtain the contribution to the gravity coefficients (Eqs.\,\ref{121105b}$-$\ref{121105f}), we average again over $\theta$ and $M$. However, since the rotation rate is now in resonance, we have that $\phi = \theta - p M $ is constant, and therefore we must retain the terms with argument $ (q = p\,k/2) $ in the expansions  (\ref{061120ga}) and (\ref{061120gb}).
The contribution to $J_2^t$ is still given by expression (\ref{121105z}), but the remaining gravity coefficients become:
\begin{eqnarray}
<C_{21}>&=& -\frac{k_f}{2} \frac{\M}{m} \left(\frac{R}{a}\right)^3 \sin \ve \left[  
\phantom{\frac{}{}} G(p/2,e) \cos \ve  \sin \phi
\right. \crm &-&
H(p/2,e) \cos^2 \frac{\ve}{2} \sin (\phi-2\lw) 
\crm &+& \left.
H(-p/2) \sin^2 \frac{\ve}{2} \sin (\phi+2\lw)
 \phantom{\frac{}{}} \right] \ , \llabel{121109c} 
\end{eqnarray}
\begin{eqnarray}
<S_{21}>&=& -\frac{k_f}{2} \frac{\M}{m} \left(\frac{R}{a}\right)^3 \sin \ve \left[  
\phantom{\frac{}{}} G(p/2,e) \cos \ve \cos \phi
\right. \crm &-& 
H(p/2,e) \cos^2 \frac{\ve}{2}  \cos (\phi-2\lw)  
\crm &+& \left.
H(-p/2,e) \sin^2 \frac{\ve}{2}  \cos (\phi+2\lw)
 \phantom{\frac{}{}} \right] \ , \llabel{121109d} 
\end{eqnarray}
\begin{eqnarray}
<C_{22}>&=& \frac{k_f}{8} \frac{\M}{m} \left(\frac{R}{a}\right)^3 \left[  
\phantom{\frac{}{}} G(p,e) \sin^2 \ve \, \cos 2 \phi
\right. \crm &+& 
2 H(p,e) \cos^4 \frac{\ve}{2} \cos 2(\phi-\lw) 
 \crm&+& \left.
2 H(-p,e) \sin^4 \frac{\ve}{2} \cos 2(\phi+\lw)
\phantom{\frac{}{}} \right] \ , \llabel{121109e} 
\end{eqnarray}
\begin{eqnarray}
<S_{22}>&=& -\frac{k_f}{8} \frac{\M}{m} \left(\frac{R}{a}\right)^3 \left[  
\phantom{\frac{}{}} G(p,e) \sin^2 \ve \,  \sin 2 \phi
\right. \crm &+&
2 H(p,e) \cos^4 \frac{\ve}{2} \sin 2 (\phi-\lw) 
 \crm &+& \left.
2 H(-p,e) \sin^4 \frac{\ve}{2} \sin 2 (\phi+\lw)
\phantom{\frac{}{}} \right] \ . \llabel{121109f} 
\end{eqnarray}
Notice that the coefficients $<C_{21}>$ and $<S_{21}>$ can only be different from zero for ``integer'' spin-orbit resonances ($1/1$, $2/1$, $3/1$, etc.), since the Hansen functions $G(p/2,e)$ and $H(p/2,e)$ are not defined when $p$ is an half-integer (Tab.\,\ref{TAB1}).

When the argument of the periapse, $\lw$, is also a fast varying angle ($\dot \lw \gg \tau_a^{-1}$), which is often the case for solid close-in planets and satellites, the resonant angle becomes $\gamma_k = \phi - k \lw$, with $k = 0$, $\pm1$, or $\pm2$ \citep{Correia_Laskar_2010}.
We can therefore also average the gravity coefficients over $\omega$, only retaining terms in  $\gamma_k$ for a given $k$ value.
For moderate obliquity, the dominating term is $\gamma_1 = \phi - \lw$.
Thus, we get
\be
<C_{21}>=<S_{21}>=0 \ , \llabel{121105x}
\ee
\begin{eqnarray}
<C_{22}>&=& \frac{k_f}{4} \frac{\M}{m} \left(\frac{R}{a}\right)^3
H(p,e) \cos^4 \frac{\ve}{2} \cos  2\gamma_1 \ , \llabel{121105v} 
\end{eqnarray}
\begin{eqnarray}
<S_{22}>&=& -\frac{k_f}{4} \frac{\M}{m} \left(\frac{R}{a}\right)^3
H(p,e) \cos^4 \frac{\ve}{2} \sin  2\gamma_1  \ . \llabel{121230a} 
\end{eqnarray}
For damped librations $\gamma_1$ is constant.
Addopting $\gamma_1 =  0$, 
i.e., the projection of the $\vi$-axis in the orbital plane points to the star at periapse,
we further get $\cos 2\gamma_1 = 1$ and $<S_{22}>=0$.

For a planet with the spin-axis normal to the orbit ($\ve = 0$), and truncating the series $H(p,e)$ to $e^2$, we retrieve the same results as in \citet{Giampieri_2004}.
In Table~\ref{tab2} we compute the gravity coefficients for the terrestrial planets and main satellites in the Solar System and compare with the observed values.
We obtain a good agreement in all situations except in the cases of Mercury, Venus and the Moon. The observed values in these three situations correspond to fossilized values acquired when their rotations were much faster than today \citep{Touma_Wisdom_1994,Correia_Laskar_2001, Correia_Laskar_2012}.

\begin{table}
\begin{center}
\caption{Low order gravity field coefficients ($\times10^{-6}$). 
\llabel{tab2} } 
\begin{tabular}{l c c c c c}
\hline
{Body} & $k_f$ & $<J_2>$ & $J_2$ & $<C_{22}>$ & $C_{22}$ \\
\hline
{Mercury}\,$^a$ & 0.928 & 0.534 & 50.3$\pm0.2$ & 0.068 & 8.1$\pm0.1$  \\
{Venus}\,$^b$ & 0.928 & 0.0516 & 4.46$\pm0.026$ & - & 0.539$\pm0.008$  \\
{Earth}\,$^b$ & 0.933 & 1063.5 & 1082.6 & - & 1.57  \\
{Mars}\,$^b$ & 1.20 & 1830.3 & 1960.5$\pm0.2$ & - & -54.73$\pm0.02$\\
{Moon}\,$^b$ & 1.44 & 8.968& 203.8 $\pm0.6$& 2.675& 22.4$\pm0.1$   \\
{Io}\,$^c$ & 1.29 &1834.5 & 1859.5$\pm2.7$& 546.8& 558.8$\pm0.8$ \\
{Europa}\,$^c$ & 1.04 & 431.0& 435.5$\pm8.2$& 129.5& 131.5$\pm2.5$ \\
{Ganymede}\,$^c$ & 0.80 & 125.97& 127.5$\pm2.9$ & 37.92& 38.26$\pm0.87$ \\
{Callisto}\,$^{c,d}$ & 1.11 &34.3 &32.7$\pm0.8$ &10.3 & 10.2$\pm0.3$\\
{Rhea}\,$^e$ & 1.25 & 780.8& 794.7$\pm89.2$& 234.9& 235.3$\pm4.76$ \\
{Titan}\,$^f$ & 1.00 & 32.9& 33.5$\pm0.6$& 9.86 & 10.02$\pm0.07$\\
\hline
\end{tabular}

The comparison is done between the values obtained through the averaged equations (Eqs.\,\ref{121102i},  \ref{121105z}, \ref{121105v}) and the values given by the observations. 
$<C_{22}>$  is computed only when the rotation is trapped in a spin-orbit resonance. 
References: $^a$ \citet{Smith_etal_2012}; $^b$ \citet{Yoder_1995}; $^c$ \citet{Schubert_etal_2004}; $^d$ \citet{Anderson_etal_2001}; $^e$ \citet{Iess_etal_2007}; $^f$ \citet{Iess_etal_2010}.
\end{center}
\end{table}

\section{Conclusion}

Using a vectorial formalism, we derived the gravity field coefficients of a planet in hydrostatic equilibrium with the tidal and centrifugal potentials.
We have made no particular assumption on the inertia tensor, so our results are valid for any rotation rate, obliquity and orbital eccentricity.
In particular, they allow us to compute the shape of the planet for an arbitrary spin-orbit resonance.

Combining expressions (\ref{121102i}),  (\ref{121105z}) and (\ref{121105v}) we get
\be
\frac{C_{22}}{J_2} = \frac{3 H(p,e) \cos^4 (\ve/2)}{(2 p)^2 + (6-9 \sin^2 \ve) (1-e^2)^{-3/2}} 
\ , \llabel{121116a}
\ee
which may considerably differ from the ratio $ 3 / 10$ observed for the satellites in the Solar System ($p=1$, $\ve = 0$, $e=0$) (Table\,\ref{tab2}).
Therefore, if we are able to measure the ratio $C_{22}/J_2$ for exoplanets in eccentric orbits, for instance, by detecting differences in the light curve at each transit \citep[e.g.][]{Barnes_Fortney_2003, Ragozzine_Wolf_2009}, we can determine in which spin-orbit the planet is locked, assuming zero obliquity.
Inversely, if we assume synchronous rotation, we can infer the obliquity of the planet.

Our work is also important for future studies on the long-term evolution of planets and satellites.
Indeed, as the shape of the planet changes from one spin-orbit resonance to another, the capture probabilities are considerably modified.
As an example, if one suppose that the Moon acquired its present figure in the past when it was closer to the Earth, we conclude that the Moon was not synchronous at the time, since $C_{22}/J_2 = 0.11$ (Tab.\,\ref{tab2}).
If the rotation was in a 3/2 spin-orbit resonance \citep{Garrick-Bethell_etal_2006}, using equation (\ref{121116a}) we can determine that $e \simeq 0.17$, 
and from expression (\ref{121105v}) that $a \simeq 25 \, R_\oplus $.

\acknowledgments

We acknowledge support from FCT-Portugal (PTDC/CTE-AST/098528/2008, SFRH/BSAB/1148/2011, PEst-C/CTM/LA0025/2011), and FAPESP (2009/16900-5, 2012/13731-0).


\end{document}